\documentclass{appolb}
\def\sun{\hbox{$\odot$}}
\def\2F1{~_2F_1}
\def\apj{ApJ\,}
\def\aap{A\&A\,}
\def\mnras{MNRAS\,}
\def\aj{AJ}
\def\cjaa{Chinese J. Astron. Astrophys.}

\def\nat{Nature\,}
\usepackage {graphicx}

\begin{document}
\title{On the product of two gamma variates with argument 2:
Application to the  luminosity function for  galaxies. %
}
\author{Lorenzo Zaninetti
\address{Dipartimento  di Fisica Generale,
 via P.Giuria 1,\\ I-10125 Turin,Italy}
}
\maketitle
\begin{abstract}
A new luminosity function for galaxies can be built starting from
the product of two random variables
X and Y represented by
a gamma variate with argument 2 .
The mean , the standard deviation and
the distribution function  of this new 
distribution are computed.
This new
probability density function 
 is assumed to  describe the mass 
distribution of galaxies.
Through   a non linear rule
of conversion from mass to luminosity 
a second new luminosity function  for galaxies is derived.
The test  of reliability of these two luminosity functions
was made on 
the Sloan Digital Sky  Survey (SDSS) in five different bands.
The  Schechter function
gives a  better fit with  respect to the 
two  new luminosity functions for galaxies here derived.
\end{abstract}

\section{Introduction}

Given  two independent non-negative random variables X and Y,
their  product XY  represents an  active field of research.
When X  and Y are  Student's $t$ random variables
the product XY is applied  in the field of finance~\cite{Su2006} .
When  X and Y
are n-Rayleigh distribution , 
the application can be the  wireless propagation
research~\cite{Salo2006}~.
In Section~\ref{secprobability}
this  paper explores
the product XY when
X and Y are  gamma variate with argument 2.
Section~\ref{vorogalaxies} explores  the  
connection between  the Voronoi Diagrams and galaxies. 
Section~\ref{sectionlum} reports two new 
luminosity functions for galaxies as deduced from the product XY.

\section{The new distribution of probability}
\label{secprobability}

The starting point is the
probability density function  ( in the following PDF)
 in length , $s$ ,
of a segment in a random fragmentation
\begin{equation}
p(s) = \lambda \exp {(-\lambda s)} ds
\quad ,
\end{equation}
where $\lambda$ is the hazard rate of the exponential  distribution.
Given  the fact that the sum , $u$ , of two  exponential
distributions has PDF 
\begin{equation}
p(u)= \lambda^2 u \exp{(- \lambda u)} du
\quad .
\end{equation}
The PDF  of the $1D$    Voronoi segments , $l$,
( the midpoint of the sum of two segments) can be found
from the previous formula  inserting  $u=2l$
\begin{equation}
p(l) = 2 \lambda l \exp{(-2 \lambda l)} d (2 \lambda l)
\quad .
\end{equation}
On transforming in normalized units $x=\frac{l}{\lambda}$
we  obtain  the following PDF 
\begin{equation}
p(x) = 2 x \exp{(-2 x) } d (2 x)
\quad .
\end{equation}
When this result is expressed as a gamma variate
we obtain the PDF (formula~(5) of  \cite{kiang})
\begin{equation}
 H (x ;c ) = \frac {c} {\Gamma (c)} (cx )^{c-1} \exp(-cx)
\quad ,
\label{kiang}
\end{equation}
where $  0 \leq x < \infty $ , $ c~>0$
and  $\Gamma (c)$ is the gamma function with argument c;
in the case of $1D$   Voronoi Diagrams $c=2$.
It was conjectured that the area in $2D$ and the volumes
in $3D$ of the   Voronoi Diagrams may be approximated
as the sum of two and three gamma variate with argument 2.
Due to the fact that the sum of n independent gamma
variates with shape parameter $c_i$ is a gamma variate with
shape parameter $c = \sum_{i}^{n} c_i$
the area and the volumes are supposed to follow a gamma variate
with argument 4 and 6 ~\cite{Feller_1971,Tribelsky2002}.
This hypothesis was later named
 "Kiang's conjecture", and equation (\ref{kiang}) was used as
a fitting function , see \cite{kumar,Zaninetti2006},
or as  an hypothesis to accept or to reject  using the standard
procedures of the data analysis,
see~\cite{Tanemura1988,Tanemura2003}.
PDF  (\ref{kiang}) can be  generalized by 
introducing  the 
dimension of the considered space,  $d(d=1,2,3)$, 
\begin{equation}
 H (x ;d  ) = \frac {2d} {\Gamma (2d)} (2dx )^{2d-1} \exp(-2dx)
\quad .
\label{kiangd}
\end{equation}

Two other PDF  are suggested for the  Voronoi Diagrams:
\begin{itemize}
\item The generalized gamma function with three parameters $(a,b,c)$
\begin{equation}
f(x;b,c,d) = c \frac {b^{a/c}} {\Gamma (a/c) } x^{a-1} \exp{(-b x^c)}
\quad ,
\end{equation}
see \cite{Tanemura2003}, \cite{Hinde1980}.  

\item A new analytical PDF  of the type
\begin{equation}
FN(x;d) =  Const \times x^{\frac {3d-1}{2} } \exp{(-(3d+1)x/2)}
\quad ,
\label{rumeni}
\end{equation}
where
\begin{equation}
Const =  
\frac 
{
\sqrt {2}\sqrt {3\,d+1}
}
{
2\,{2}^{3/2\,d} \left( 3\,d+1 \right) ^{-3/2\,d}\Gamma  \left( 3/2\,d+
1/2 \right) 
}
\quad ,
\end{equation}
and $d(d=1,2,3)$ represents the 
dimension of the considered space ,  see \cite{Ferenc_2007}.
\end{itemize}

Experimentally determined physical quantities are usually derived from
combinations of measurements, each of which may be considered a random
variable subject to a known distribution law. 
The distribution law of the
sought-for physical quantity, however, is generally not known or
determinable analytically, except for a linear superposition of random
variables , i.e. the sum of n independent gamma
variates, \cite{Silverman2003}.
Physical quantities represented as products  of
random variables are of especial interest.
Computer simulations of products  of normally
distributed random variables , as an example ,
lead to distributions that are not normal, in
some cases markedly so with pronounced skewness, depending upon the
parameters of the component distributions, \cite{Silverman2003}.
Consider , for example,   the  product  of  two random variables 
$X\approx N(0,1)$ and the random variable  $Y\approx N(0,1)$,
the PDF of $V=XY$ is  , see \cite{Glen2004} ,  
\begin {equation}
  h(v) = \left  \{ \begin {array} {ll}
           K_0(v * signum (v))/ \pi   &  -\infty < v < 0   \\
           K_0(v * signum (v))/ \pi   &       0  < v < \infty     ~.
\end {array}
    \right .
\end {equation}
This PDF  has a pole  at  $v=0$ .

We now explore the product
 of two  gamma variate with argument 2.
We recall that if
 $X$ is  a random variable of the continuous type with 
 PDF 
, $f(x)$,
 which
is defined  and positive on the interval
 $  0 \leq x < \infty $
and   similarly  if
$Y$
is  a random variable of the continuous type with PDF
$g(y)$
 which is defined
and
positive $  0 \leq y < \infty $ ,  the PDF  of $V = XY$ is
\begin{equation}
h(v) =
\int_0^\infty g (\frac{v}{x}) f(x) \frac{1}{x} dx
\quad .
\end{equation}
Here the case of equal limits of integration will
be explored , when this is not true  difficulties arise
\cite{Springer1979,Glen2004} .
When $f(x)$  and   $g(y)$ are
 gamma variates with argument 2
the PDF is
\begin{equation}
h(v) =
\int_0^\infty
\frac {
16\,{e^{-2\,x}}v{e^{-2\,{\frac {v}{x}}}}
}
{
x
}
dx
=
32\,v{\it K_0} \left( 4\,\sqrt {v} \right)
\quad ,
\label{bessel}
\end{equation}
where $ K_{\nu}(z) $ is the modified Bessel function of
the second kind \cite{Abramowitz1965,press}
with $\nu$ representing the order, in our case 0.
The distribution function ( in the following DF) is
\begin{equation}
16\,{v}^{2}{\it K_0} \left( 4\,\sqrt {v} \right) {\it \2F1}
 \left( 1,2;3;4\,v \right) +16\,{v}^{5/2}{\it K_1} \left( 4
\,\sqrt {v} \right) {\it \2F1} \left( 1,3;3;4\,v \right) 
\quad ,
\end{equation}
where ${\2F1(a,b;\,c;\,v)}$ is a regularized hypergeometric function,
see~\cite{Abramowitz1965,Seggern1992,Thompson1997}.

\begin{figure}
\begin{center}
\includegraphics[width=10cm]{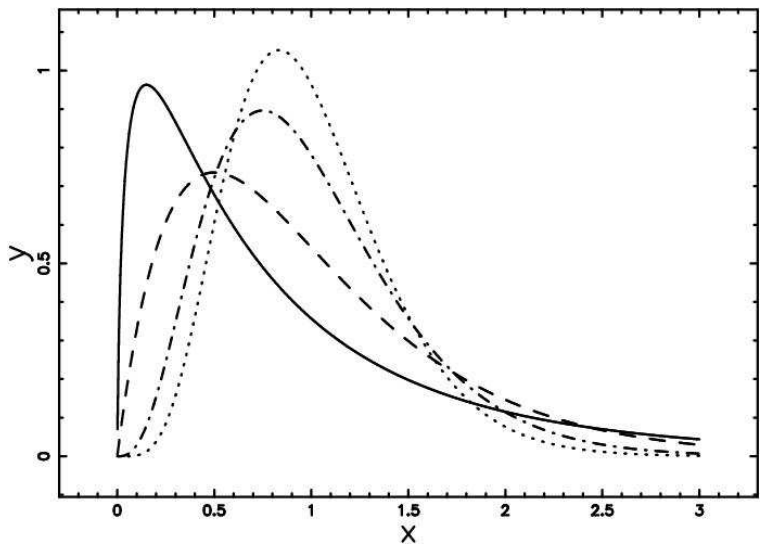}
\end  {center}
\caption {
Plot of  $h(x)$ as function of  $x$
 (full line) ,
 H (x ;c )  when $c=2$  (dashed),
 H (x ;c )  when $c=4$  (dot-dash-dot-dash)  and
 H (x ;c )  when $c=6$  (dotted).
}
\label{besselk}
\end{figure}

The mean of the  new  PDF, $h(v)$, as represented
by formula~(\ref{bessel}) is
\begin{equation}
<v> =\int_0^\infty
v \times 32\,v{\it K_0} \left( 4\,\sqrt {v} \right)  dv  = 1
\quad ,
\end{equation}
and the variance
\begin{equation}
\sigma^2  =\int_0^\infty
(v-1)^2 \times 32\,v{\it K_0} \left( 4\,\sqrt {v} \right)  dv  =
\frac {5} {4}
\quad .
\end{equation}
The mode , $m$ ,  is at  v =0.15067~.
Figure~\ref{besselk} reports our function $h(x)$ as well 
the Kiang function $H (x ;c ) $ for three values of $c$.
Asymptotic series are
\begin{equation}
h(v) \sim
-16\, \left( 2\,\ln  \left( 2 \right) +\ln  \left( v \right) +2\,
\gamma \right) v
\quad   v \ll 1
\quad ,
\end{equation}
\begin{equation}
h(v) \sim - \frac{ \sqrt {2}\sqrt {\pi }{e^{-{\frac {4}{\sqrt {
\frac{1}{v}}}}}} \left( - 32+\sqrt {\frac{1}{v}} \right) } { 4\,
\left( \frac{1}{v} \right) ^{3/4} } \quad   v \gg 1 \quad ,
\end{equation}
where $\gamma$ is the Euler-Mascheroni constant.

\section{Voronoi Diagrams and galaxies}
\label{vorogalaxies}
The applications of the Voronoi Diagrams,  see \cite{voronoi_1907}
and   \cite{voronoi},  in
astrophysics started with ~\cite{kiang} where 
through a Monte Carlo experiment,   the area distribution in 2D and volume
distribution in 3D were deduced. 
The  application of  the Voronoi Diagrams to the distribution of
galaxies started with~\cite{icke}, where a sequential clustering
process was adopted in order to insert the initial seeds. 
Later
a  general algorithm for simulating
one-dimensional lines of sight through a cellular universe
was introduced \cite{pierre}. 
The
large microwave background temperature anisotropies over angular
scales up to one degree were calculated using a Voronoi model for
large-scale structure formation in \cite{barrow} and
\cite{coles1991}.
The intersections
between lines that represent 
the 'pencil beam' surveys
and  the faces of 
a three-dimensional Voronoi tessellation 
has been investigated by 
\cite{Ikeuchi1991}  where 
an exact expression is derived for the distribution of spacings of these
intersections.
Two  algorithms (among others) that allow to   
detect structures from galaxy
positions and magnitudes are briefly reviewed   : 
\begin{itemize}
\item
 A Voronoi Galaxy Cluster Finder (VGCF)
that uses galaxy positions and magnitudes to find clusters 
and determine their
main features: size, richness and contrast above the background , 
see~\cite{Ramella,Ramella1999}.
\item 
An automated procedure for structure finding, involving the Voronoi
tessellation
allows to build  a catalogue (called PF) of galaxy structures 
(groups and clusters)  in
an area of 5,000 square degrees in the southern hemisphere,
see~\cite{Panko2006a,Panko2006b}.
\end{itemize}

This section first explores  how the fragmentation of a 2D  layer 
as due to the 2D Voronoi Diagrams can be useful or not
in describing the mass distribution of galaxies.
The large scale structures  of our universe 
are then explained  by  the 3D Voronoi Diagrams in the
second section.

\subsection{Mass distribution}

Here we  Analise  the fragmentation of a 2D
layer of thickness which is  negligible with respect
to the main dimension.
A typical dimension of the layer can be found  as follows.
The averaged observed diameter of the galaxies is
\begin{equation}
\overline{D^{obs}} \approx  0.6  {D_{max}^{obs}} = 2700 \frac{Km}{sec}
= 27~Mpc
\quad ,
\label{dobserved}
\end {equation}
where $D_{max}^{obs}=4500~\frac{Km}{sec}$  corresponds to the
extension of the maximum void visible on the CFA2 slices. 
In the
framework of the theory of the primordial explosions
,see~\cite{charlton1986} and \cite{ferraro}, this means that the
mean observed area of a bubble ,$\overline{A^{obs}} $, is
\begin{equation}
\overline{A^{obs}} \approx  4 \pi (\frac{D_{max}^{obs}}{2})^2
=2290 Mpc^2
\quad .
\label{aobserved}
\end {equation}
The averaged area of a face of a  Voronoi polyhedron ,
$\overline{A_V} $, is
\begin{equation}
\overline{A_V}  = \frac{\overline{A^{obs}}} {\overline{N_F}}
\quad ,
\end {equation}
where  ${\overline{N_F}}$ is the averaged number 
of irregular
faces of the Voronoi polyhedron,
i.e.  ${\overline{N_F}}$=16 , see~\cite{okabe,Zaninetti2006}.
The averaged side of a face of a irregular polyhedron , $L_V$ ,
is
\begin{equation}
\overline{L_V}  \approx  \sqrt {\overline{A^{obs}}} \approx 12~Mpc
\quad .
\end {equation}
The thickness of the layer  , $\delta$ ,
can be derived from the shock theory
, see~\cite{deeming}, and is  1/12 of the radius
of the advancing shock ,
 \begin{equation}
\delta = \frac { {D_{max}^{obs}}  } {2 \times 12 } \approx 1.12 Mpc
\quad .
\end{equation}
The number of galaxies in this typical layer , $N_G$, can be found
by 
multiplying $n_*\approx 0.1$ , the density of galaxies , by the
volume of the cube of side 12~$Mpc$ : i.e. $N_G \approx 172 $.

\begin{figure}
\begin{center}
\includegraphics[width=10cm]{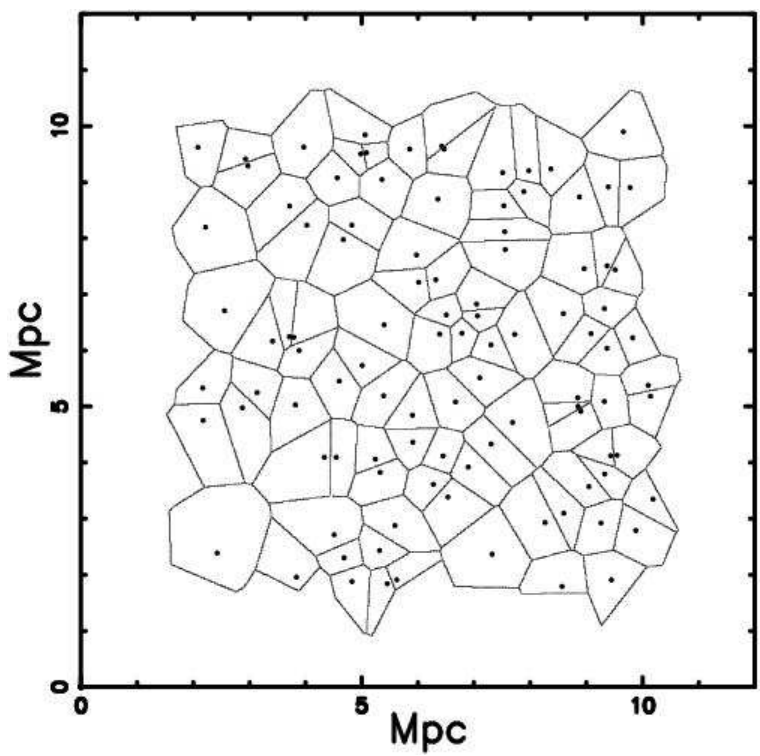}
\end  {center}
\caption {
The   Voronoi--Diagram in 2D when random seeds are used.
The selected region  comprises  102 seeds.
}
\label{area2D_random}
\end{figure}

A first application 
of the  new  PDF, $h(v)$, as represented
by formula~(\ref{bessel}) , 
can be a test on the area distribution
of the   Voronoi polygons , 
see Figure~\ref{area2D_random}.
Does the area distribution
of the irregular polygons follow the
sum or the product of two gamma variates
with argument 2 ?.
In order to answer  this question we fitted the sample
of the area
with $h(v)$ , the new PDF ,
with a gamma variate with argument
4
and  with a gamma variate with the  argument
as deduced from the sample.
The results are reported  in Table~\ref{data}.
 \begin{table}
 \caption[]{The $\chi^2$  of data fit when the number
            of classes is 10 for three PDF }
 \label{data}
 \[
 \begin{array}{lc}
 \hline
PDF    &   \chi^2  \\ \noalign{\smallskip}
 \hline
 \noalign{\smallskip}
h(x)                       &    62.8                 \\ \noalign{\smallskip}
H (x ;c )  ~when~ c=4      &    28.3                  \\ \noalign{\smallskip}
H (x ;c )  ~when~ c=3.7    &    23.55                \\ \noalign{\smallskip}
FN(x;d  )~ Ferenc ~\&~ Neda~ formula~(\ref{rumeni}) ~when~ d=2   &    20.2                  \\ \noalign{\smallskip}
 \hline
 \hline
 \end{array}
 \]
 \end {table}
From a careful inspection of Table~\ref{data} it is possible to conclude
that the area distribution of the irregular     Voronoi polygons
is better described by the sum of two gamma variates
with argument 2 rather than by the product.

\subsection{Spatial  dependence}

\begin{figure}
\begin{center}
\includegraphics[width=10cm]{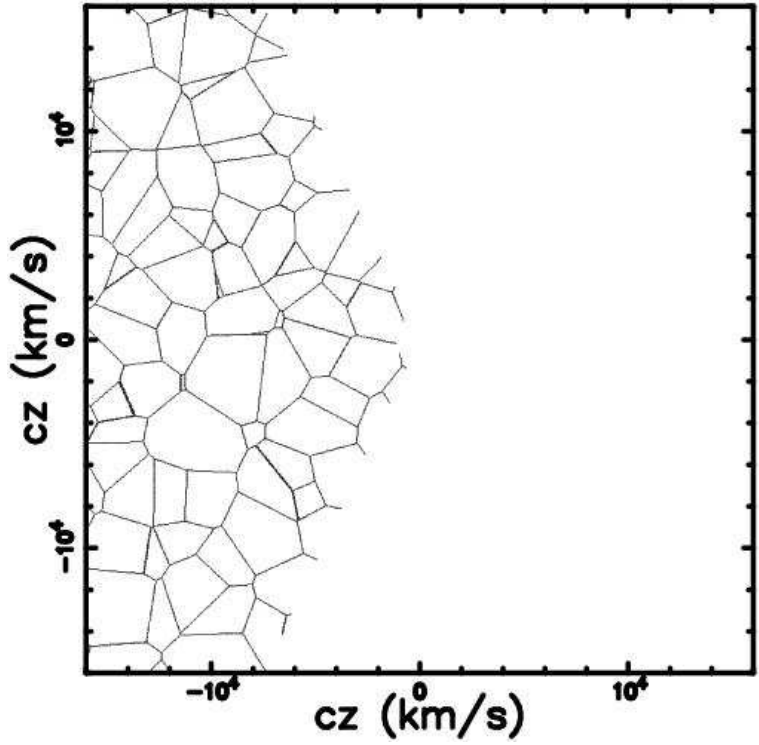}
\end {center}
\caption {
Portion of the Voronoi--Diagram $V_p(2,3)$ when random seeds are used;
cut on the  x--y plane ~.
For astronomical  purposes we only report a slice
 $130^{\circ}$ long.
The  parameters
are {\it pixels}~= 800
       , {\it N}~= 1900
   and   {\it side}~= 2 $\times$ 16000 Km/sec
          \label{taglio_mezzo}%
 }  
    \end{figure}

Our  method considers a  3D lattice
made of   ${\it pixels}^{3}$ points~:
present in this lattice are $N_s$ seeds generated
according to a random process.
All the computations are usually performed on this mathematical
lattice; 
the conversion to the physical lattice
is obtained   by multiplying the unit
by $\delta=\frac{side}{pixels -1}$ , where {\it side}
is the  length of the square/cube expressed in the physical
unit  adopted.
The tessellation in $\Re^3$ is firstly analyzed
through a planar section .
Given a section  of the cube
(characterized , for example, by $k=\frac{pixel}{2}$)
the  various $V_i$ (the volume belonging
to the seed $i$)
 may or may not cross the  little
cubes belonging to the two dimensional lattice .
Following the nomenclature introduced by~\cite{okabe} we  call
the intersection between a plane and the cube previously described
as  $V_p(2,3)$; a typical  example  is shown in 
Figure~\ref{taglio_mezzo}.

\begin{figure}
\begin{center}
\includegraphics[width=10cm]{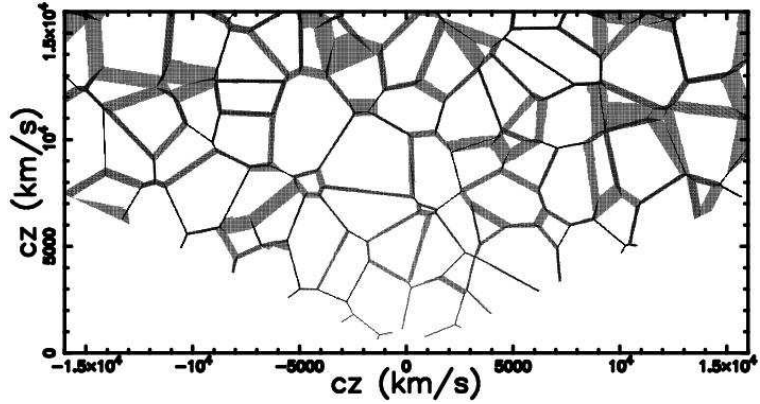}
\end {center}
\caption {
Polar plot
of the  little cubes belonging to a
slice   $130^{\circ}$~long  and $6^{\circ}$
wide.
Parameters  as in Figure~\ref{taglio_mezzo}.}
          \label{cfaslices}%
    \end{figure}

For astronomical purposes is also  interesting to plot 
the  little cubes belonging to a slice 
 of $6^{\circ}$
wide and about $130^{\circ}$ long, see Figure~\ref{cfaslices}.

\section{A new luminosity function for galaxies}
\label{sectionlum}
The new  PDF as  given by formula~(\ref{bessel})
can represent a   PDF  in mass for the galaxies.
The luminosity function , in the following LF, 
for galaxies is then 
deduced by introducing a linear and a non linear 
relationship between mass and luminosity.
A special section is devoted to the parameters 
determination of the two new LF  for galaxies.
The dependence of the number of galaxies with the redshift 
is then analyzed by  adopting the first new LF.

\subsection{Schechter~luminosity~function}

A model for the LF  of galaxies 
is  the Schechter function
\begin{equation}
\Phi (L) dL  = (\frac {\Phi^*}{L^*}) (\frac {L}{L^*})^{\alpha}
\exp \bigl ( {-  \frac {L}{L^*}} \bigr ) dL \quad  ,
\label{equation_schechter}
\end {equation}
where $\alpha$ sets the slope for low values of $L$ , $L^*$ is the
characteristic luminosity and $\Phi^*$ is the normalization. 
This
function  was suggested  by \cite{schechter} in order to
substitute other analytical expressions, see  for example, 
formula~(3) in~\cite{kiang1961}. 
Other  interesting quantities are  the  mean
luminosity per unit volume,
$j$ ,
\begin{equation}
\label{eqnj}
j = \int ^{\infty}_0 L \Phi(L) dl = 
{\it L^*}\,{\it \Phi^* }\,\Gamma  \left( \alpha+2 \right) 
\quad ,
\end{equation}
and the averaged luminosity ,${ \langle L \rangle }$ ,
\begin{equation}
\label{lmedia}
\langle L \rangle =\frac{j}{\Phi^*}= 
{\it L^*}\,\Gamma  \left( \alpha+2 \right) 
 \quad,
\end{equation}
where  $\Gamma$ is the gamma function and 
its  appearance  is explained in~\cite{Efstathiou1988}.
An astronomical form of
equation~(\ref{equation_schechter}) can be deduced by introducing the
distribution in absolute magnitude
\begin{eqnarray}
\Phi (M)   dM=&&(0.4  ln 10) \Phi^* 10^{0.4(\alpha +1 ) (M^*-M)}\nonumber\\
&& \times \exp \bigl ({- 10^{0.4(M^*-M)}} \bigr)  dM \quad  ,
\label {equation_schechter_M}
\end {eqnarray}
where $M^*$ is the characteristic magnitude as derived from the
data. 
At present this function  is widely  used and  
Table~\ref{parameters} reports the parameters 
from  the following catalogs  
\begin{itemize}
\item 
The 2dF Galaxy
Redshift Survey (2dFGRS) based on a preliminary subsample of 45000 galaxies,
see~\cite{Cross2001}.

\item  The $r^\ast$-band LF  for a sample of 11,275
       galaxies from the  Sloan Digital Sky
       Survey (SDSS) , see~\cite{blanton}.
\item  The  galaxy LF
        for a sample of  10095 galaxies 
        from the Millennium Galaxy Catalogue (MGC)
        , see~\cite{Driver2005}.  
\item
The CFA Redshift Survey \cite{Marzke_1994b} that  covered 9063
galaxies with Zwicky $m$ magnitude $<$ 15.5 
to calculate the galaxy LF  over the range 13 $<$ M $<$ 22. 
\end{itemize}
\begin{table} 
 \caption[]{The parameters of the Schechter function  from 
     2dFGRS , SDSS , MGC and CFA . } 
 \label{parameters} 
 \[ 
 \begin{array}{lcccc} 
 \hline 
 \hline 
 \noalign{\smallskip} 
parameter & 2dFGRS & SDSS~(r^*)~ band  & MGC & CFA  \\ \noalign{\smallskip}
M^* ~ [mags]      &   -19.75 \pm 0.05   & -20.83 \pm 0.03 &  -19.60 \pm 0.04  &
-18.79 \pm 0.1\\ \noalign{\smallskip}
\alpha    &   -1.09  \pm 0.03    & -1.2  \pm 0.03 &  -1.13 \pm 0.02 
&  -1. \pm 0.1 
\\ \noalign{\smallskip}
\Phi^* ~[h~Mpc^{-3}]    &  (2.02   \pm 0.02)10^{-2}     &(1.46   \pm 0.12)10^{-2}      
&  (1.77   \pm 0.15)10^{-2} 
&  (4.0   \pm 0.1)10^{-2} 
\\ \noalign{\smallskip}
 \hline 
 \hline 
 \end{array} 
 \] 
 \end {table}
Over the years many modifications  have been  made  
to the standard Schechter function in order   to improve its fit:
we report three of them.
When the fit  of the rich clusters LF 
is not satisfactory
a two-component Schechter-like function is introduced
, see~\cite{Driver1996} 
\begin{eqnarray}
L_{max} > L > L_{Dwarf}  : \quad 
\Phi (L) dL  = (\frac {\Phi^*}{L^*}) (\frac {L}{L^*})^{\alpha}
\exp \bigl ( {-  \frac {L}{L^*}} \bigr ) dL \quad  ,
\nonumber  \\
 \\
L_{Dwarf} > L > L_{min}  : \quad 
\Phi (L) dL  = (\frac {\Phi_{Dwarf}}{L^*}) (\frac {L}{L_{Dwarf}})^{\alpha_{
Dwarf}}
 dL \quad  ,
\nonumber 
\end {eqnarray}
where 
\begin{eqnarray}
\Phi_{Dwarf}   = \Phi^* (\frac {L_{Dwarf} }{L^*})^{\alpha}
\exp \bigl ( {-  \frac {L_{Dwarf} }{L^*}} \bigr ) 
\quad .
\nonumber 
\end{eqnarray}
This two-component function defined  between
$L_{max}$ and $L_{min}$ 
has  two additional parameters:
$L_{Dwarf}$ which  represents  the magnitude where 
dwarfs first dominate over giants and  ${\alpha_{Dwarf}}$ 
the faint slope  parameter for the dwarf population.

Another function  introduced in order to 
fit  the case
of extremely low luminosity  galaxies 
is the  double  Schechter function  , see~\cite{Blanton_2005} :
\begin{equation}
\Phi(L) dL = \frac{dL}{L_\ast}
  \exp(-L/L_{\ast}) \left[
\phi_{\ast,1}
\left( \frac{L}{L_{\ast}} \right)^{\alpha_1} +
\phi_{\ast,2}
\left( \frac{L}{L_{\ast}} \right)^{\alpha_2}
  \right]
\quad ,
\end{equation}
where the parameters $\Phi^*$ and $\alpha$ which  characterize 
the    Schechter function have  been doubled in $\phi_{\ast,1}$
and  $\phi_{\ast,2}$.

\subsection{A linear mass-luminosity relationship}

We start by assuming that the  mass of the galaxies , ${\mathcal M}$,
is  distributed as
$h(\mathcal M)$.
We then assume a linear relationship between mass of galaxy
and luminosity , $L$ ,   
\begin{equation}
L   =  {\mathcal R} { \mathcal M}
\quad  ,
\label{linear}
\end{equation}
where $\mathcal R$ represents the mass luminosity ratio 
 $\approx (10-15)$ , see  \cite{Padmanabhan_III_2002}.
When   ${ L^*}$ represents the scale of the luminosity.
Equation~(\ref{bessel}) changes to 
\begin{equation}
\Psi ({L }) d{L }
      = \Psi^* \times
\frac{
32\,L {\it K_0} \left( 4\,{\frac {\sqrt {L}}{\sqrt {L^*}}} \right) 
}
{
{L^*} 
}
d\frac  {L }{L^*}
\label{psilinear} \quad.
\end {equation}
where $\Psi^*$  is  a normalization factor which defines the
overall density of galaxies , a  number  per cubic $Mpc$.
The mathematical range of existence is  $  0 \leq L < \infty $.
The  mean
luminosity per unit volume,
$j$ ,
\begin{equation}
j = \int ^{\infty}_0 L \Psi(L) dl =
  L^* \Psi^*
\quad .
\label{jtotlinear}
\end{equation}

The relationship connecting the absolute magnitude, $M$ ,
 of a
galaxy  with its luminosity is
\begin{equation}
\frac {L}{L_{\sun}} =
10^{0.4(M_{bol,\sun} - M)}
\quad ,
\label{mlrelation}
\end {equation}
where $M_{bol,\sun}$ is the bolometric luminosity
of the sun , which   according to \cite{cox}
is $M_{bol,\sun}$=4.74~.

A more convenient form in terms of  the absolute
magnitude $M$ is 
\begin{eqnarray}
\Psi (M) dM  =
12.8\,{\it \Psi^*}\,{10}^{ 0.8\,{\it M^*}-
0.8\,{\it M}}{\it K_0} \left(  4.0\,{10}^{
 0.2\,{\it M^*}- 0.2\,{\it M}} \right) \ln 
 \left( 10 \right) 
  dM \quad.
\label{equation_mia_linear}
\end {eqnarray}
This data oriented function contains the  parameters $M^*$ 
and  $\Psi^*$ which  can be derived from the operation of fitting
the observational data.

\begin{figure}
   \centering
\includegraphics[width=10cm]{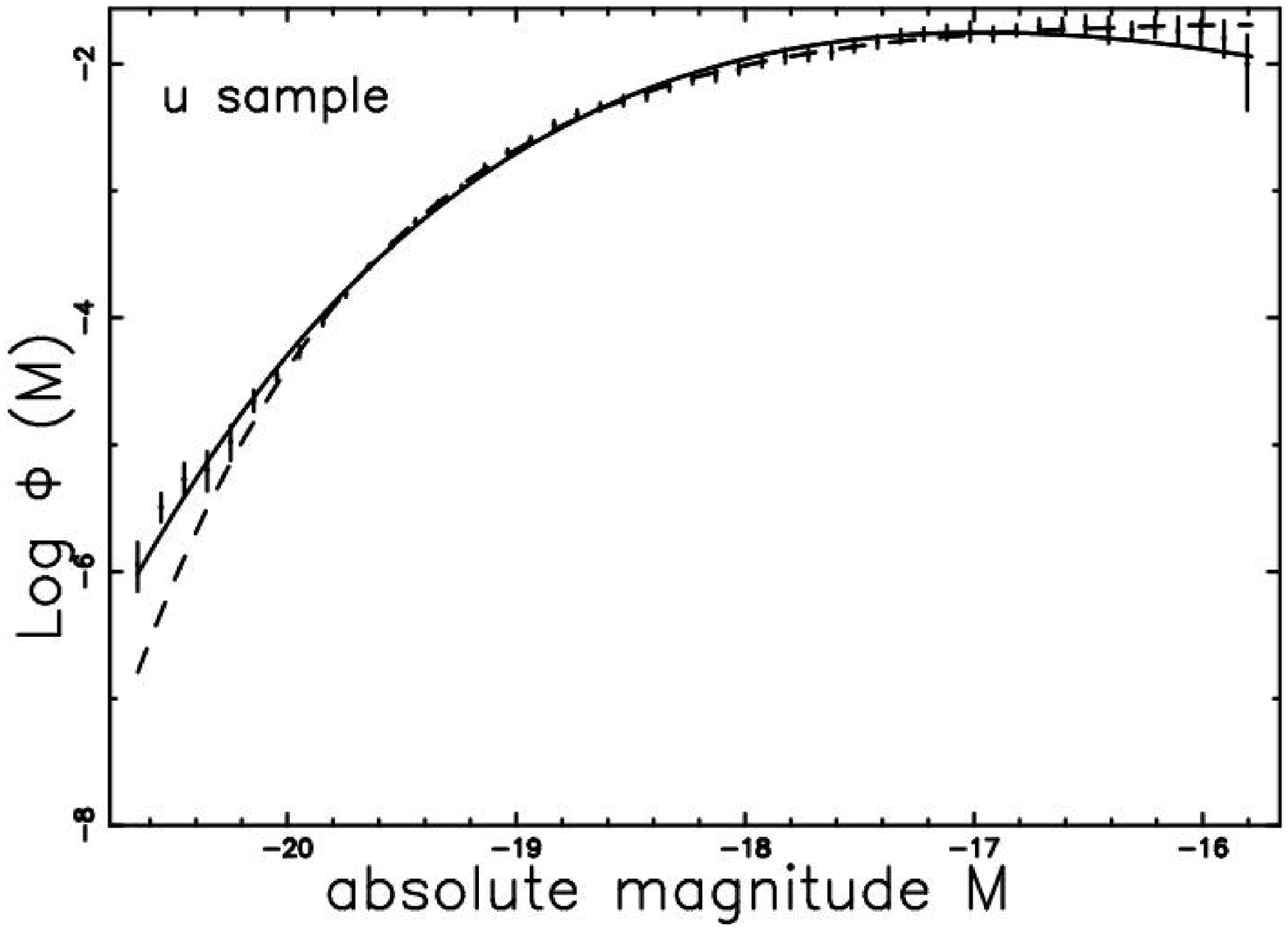}
\caption {The LF  data of  
SDSS($u^*$)  are represented through the error bar. 
The fitting continuous  line represents  our   LF
(\ref{equation_mia_linear})  
and the dotted 
line   represents the Schechter function.
 }
          \label{due_u} 
    \end{figure}

\begin{figure}
   \centering
\includegraphics[width=10cm]{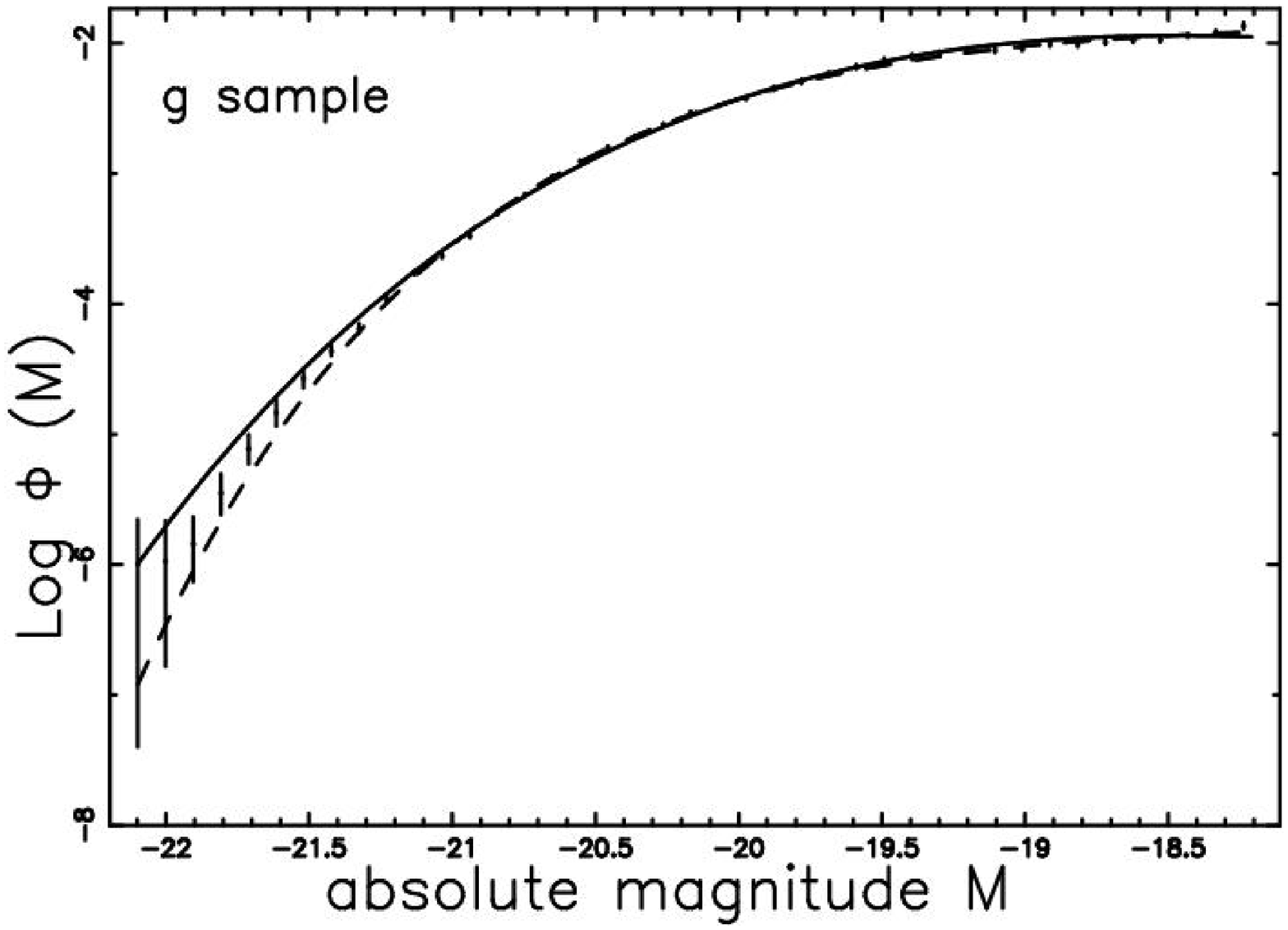}
\caption {The LF  data of  
SDSS($g^*$)  are represented through the error bar. 
The fitting continuous  line represents  our   luminosity function
(\ref{equation_mia_linear})
and the dotted 
line   represents the Schechter function.
 }
          \label{due_g} 
    \end{figure}

\begin{figure}
   \centering
\includegraphics[width=10cm]{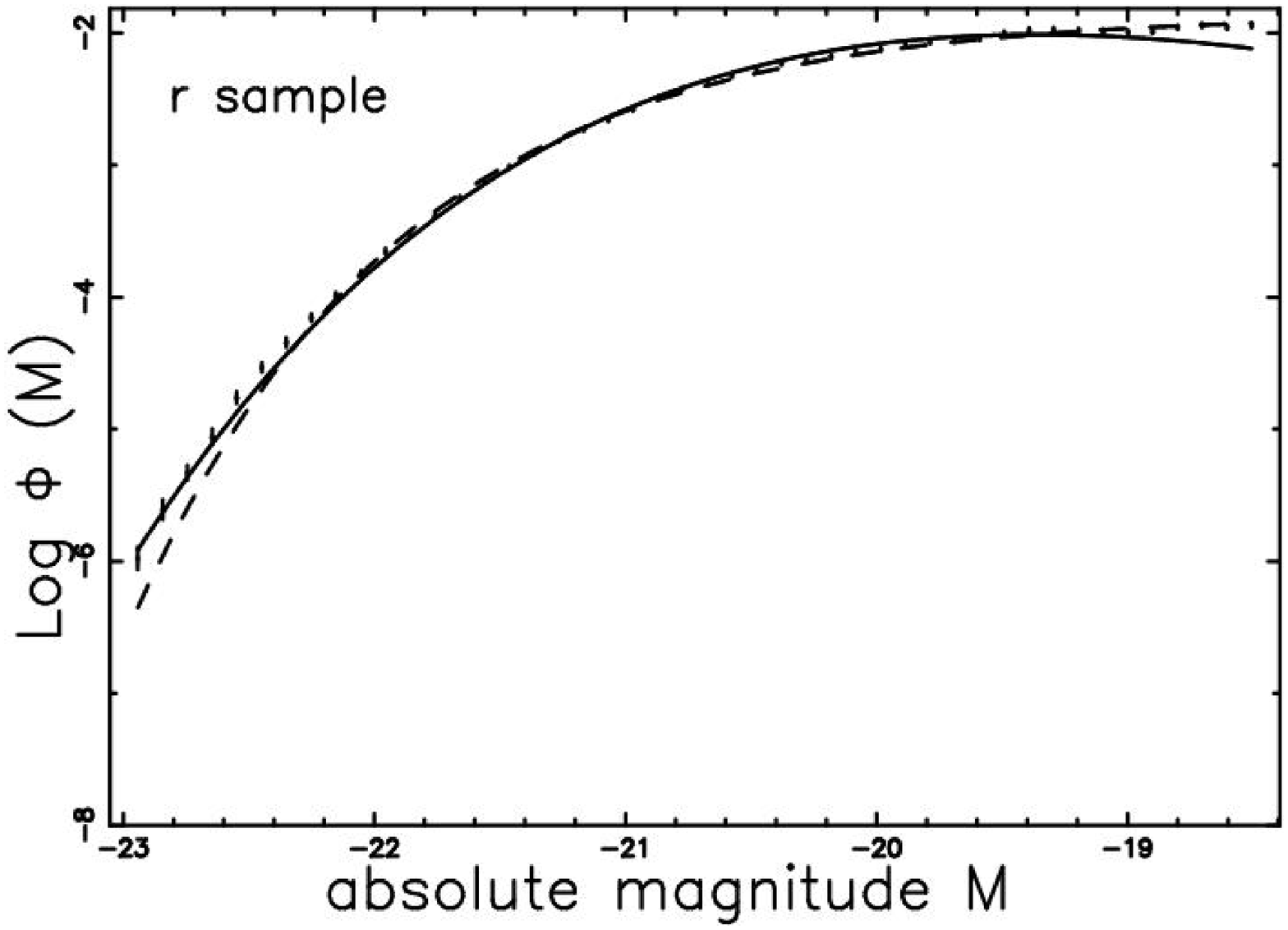}
\caption {The LF  data of  
SDSS($r^*$)  are represented through the error bar. 
The fitting continuous  line represents  our   LF
(\ref{equation_mia_linear})
and the dotted 
line   represents the Schechter LF.
 }
          \label{due_r} 
    \end{figure}

\begin{figure}
   \centering
\includegraphics[width=10cm]{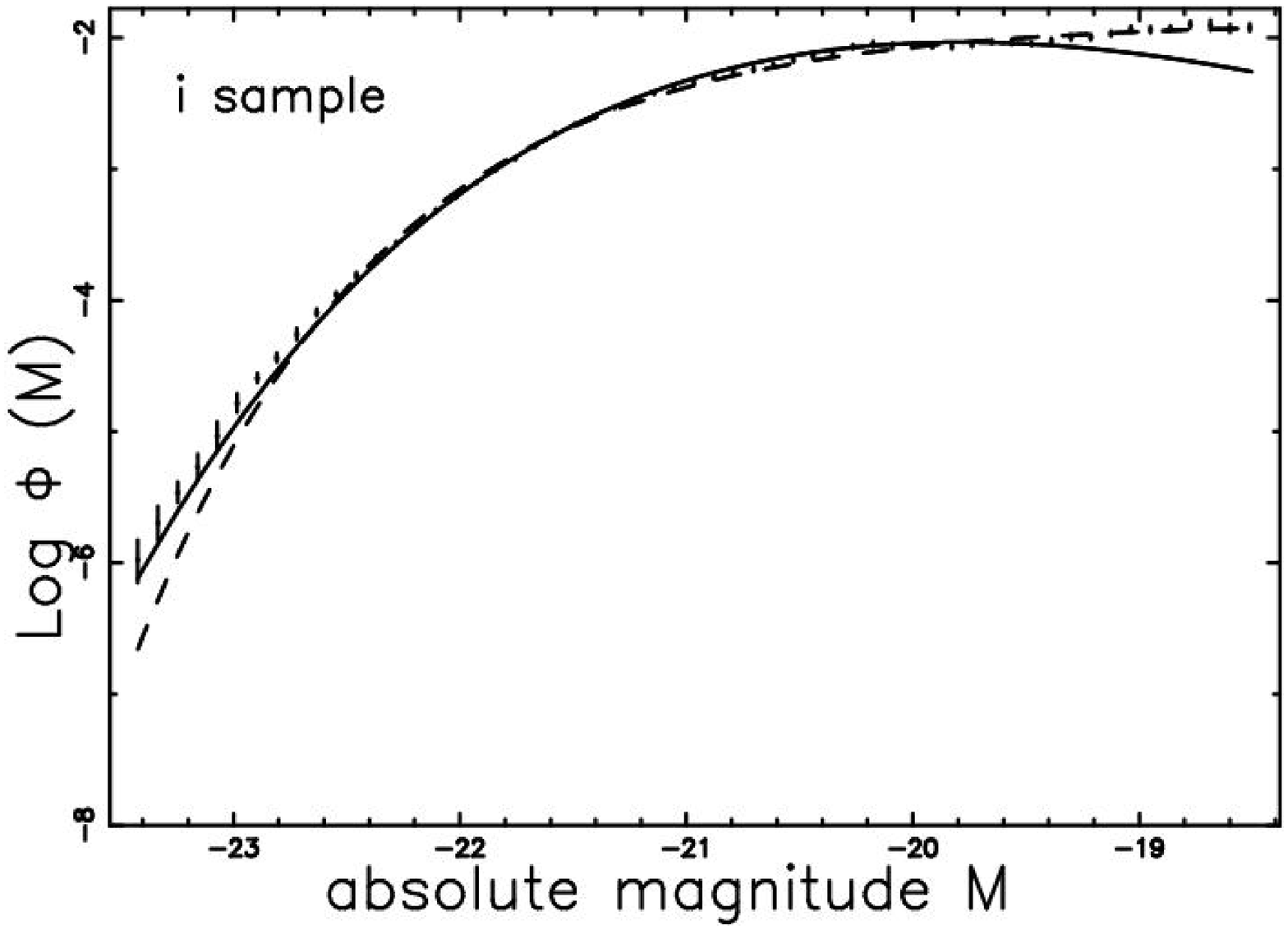}
\caption {The LF  data of  
SDSS($i^*$)  are represented through the error bar. 
The fitting continuous  line represents  our   LF
(\ref{equation_mia_linear})
and the dotted 
line   represents the Schechter LF.
 }
          \label{due_i} 
    \end{figure}

\begin{figure}
   \centering
\includegraphics[width=10cm]{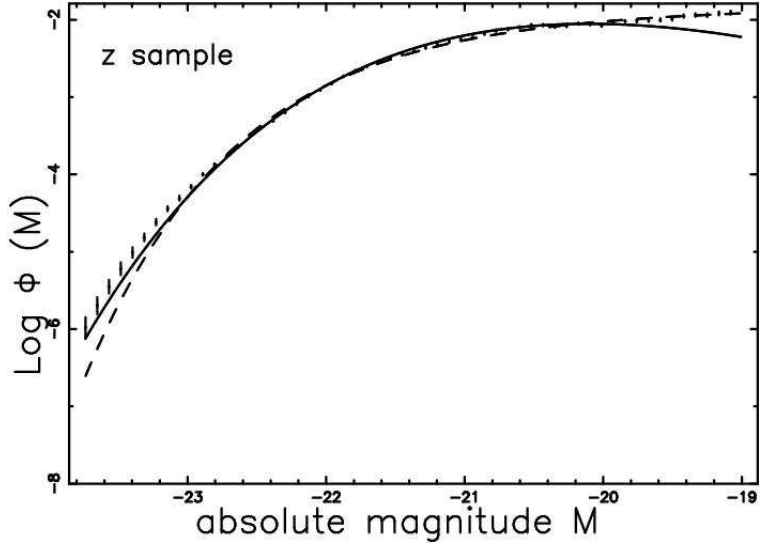}
\caption {The LF   data of  
SDSS($z^*$)  are represented through the error bar. 
The fitting continuous  line represents  our   luminosity function
(\ref{equation_mia_linear})
and the dotted 
line   represents the Schechter LF.
 }
          \label{due_z} 
    \end{figure}

In order to make a comparison between our LF and  the Schechter LF we first down-loaded 
the data of the LF for galaxies 
in the five bands of   SDSS  available at  
http://cosmo.nyu.edu/blanton/lf.html.
The LF for galaxies as obtained from the astronomical observations 
ranges in magnitude from a minimum value , $M_{min}$ , 
to a maximum value , $M_{max}$ ; details can be found in  
\cite{lin} and \cite{Machalski2000}.
For our purposes 
we then introduced an upper limit , $M_{lim}$ , 
for the absolute magnitude in order to check 
the range  of reliability of our LF 
as represented by equation~(\ref{equation_mia_linear}).
It is interesting to stress that $M_{lim}$ is used 
only for internal reasons and is connected to how
the LF,  as a function of the absolute magnitude.
reaches the maximum.
Table~\ref{datamia} reports the original range in magnitude 
of the astronomical data ,
the selected range adopted for testing purposes ,
the three parameters of our function ,
the $\chi^2$ of the fit and the 
$\chi^2$  of the  Schechter function
for the five bands of SDSS.
 \begin{table}
 \caption[]{
The full range in magnitudes , 
the selected range in magnitudes ,
the parameters of  our  function
(\ref{equation_mia_linear})  ,
$\chi^2$ ,  $AIC$ and  $BIC$                      
  of our function and the Schechter function 
for the SDSS catalog 
}
 \label{datamia}
 \[
 \begin{array}{lccccc}
 \hline
parameter    &  u^*   &  g^* & r^*  & i & z  \\ \noalign{\smallskip}
 \hline
 \noalign{\smallskip}
full~range~[mags]          
&-22   , -15.8       
&-23.4 , -16.3            
&-24.48 , -16.3            
&-24.5 , -17.2            
&-23.7 , -17.4            
\\ \noalign{\smallskip}
selected~range~[mags]      
&-20.65   ,  -15.8
&-22.09,  -18.2                      
&-22.94 ,  -18.5                      
&-23.42 ,  -18.5
&-23.7,   -19
\\ \noalign{\smallskip}
M^* [mags]                  
&  -17.23
&  -18.74
&  -19.63
&  -20.05
&  -20.37              
\\ \noalign{\smallskip}
\Psi^*                      
&   0.052                
&   0.033 
&   0.028
&   0.027
&   0.026               
\\ \noalign{\smallskip}
\chi^2                      
&   563 
&   1151
&   2758
&   4202 
&   4588          
\\ \noalign{\smallskip}
AIC, k=2                       
&    567
&    1155  
&    2762
&    4206
&    4592        
\\ \noalign{\smallskip}
BIC, k=2                       
&    575
&    1163
&    2770 
&    4215
&    4601        
\\ \noalign{\smallskip}
\chi^2~Schechter
&   330
&   456
&   1497
&   1916
&   2694     
\\ \noalign{\smallskip}
AIC~Schechter , k=3                       
&   336
&   462
&   1503
&   1922
&   2700     
\\ \noalign{\smallskip}
BIC~Schechter, k=3                       
&   349
&   474
&   1515
&   1935
&   2713     
\\ \noalign{\smallskip}
 \hline
 \hline
 \end{array}
 \]
 \end {table}
The Schechter function , the new  function and the data are
reported in 
Figure~\ref{due_u} ,
Figure~\ref{due_g} ,
Figure~\ref{due_r} ,
Figure~\ref{due_i} ,
and Figure~\ref{due_z}
when the $u^*$,$g^*$ ,$r^*$ , $i^*$  and $z^*$ 
bands of SDSS are  considered.


\subsection{A non linear mass-luminosity relationship}

Also here  we  assume that the  mass of the galaxies , ${\mathcal M}$,
is  distributed as
$h(\mathcal M)$.
The first   transformation  is
\begin{equation}
\mathcal M   = \left ( \frac  {L  }{L^*}
\right ) ^{\frac{1}{a} } \quad  ,
\label{massa}
\end{equation}
where  $L$  is the luminosity ,
$1/a$   is  an exponent  that connects the mass to the
luminosity  and  ${ L^*}$ represents the scale of the luminosity.
Equation~(\ref{bessel}) changes to
\begin{equation}
\Psi^{NL} ({L }) d{L }
      =
\frac {\Psi^*}{a} \,32\,{L}^{-{\frac {-2+a}{a}}}{L^*}^{-\frac{2}{a}   }{\it
K_0} \left( 4
\,{L}^{ \frac {1} {2a} }{L^*}^{-\frac {1} {2a}   } \right)
      d\frac  {L }{L^*}
\label{psi} \quad , 
\end {equation}
where $\Psi^*$  is  a normalization factor  and the apex {NL} stands
for nonlinear.
The  mean
luminosity per unit volume,
$j$ ,
\begin{equation}
j = \int ^{\infty}_0 L \Psi^{NL}(L) dl =
 {4}^{-a} L^* \Psi^* \left( \Gamma  \left( 2+a \right)  \right) ^{2}
\quad .
\label{jtot}
\end{equation}
The second
transformation connects the luminosity with the absolute
magnitude
\begin{eqnarray}
\Psi^{NL} (M) dM  =
12.8\,{\it \Psi^*}\,{10}^{- 0.8\,{\frac {- {\it
M^*}+{\it M}}{a}}}{\it K_0} \left(  4.0\,{10}^{-
 0.2\,{\frac {- {\it M^*}+{\it M}}{a}}} \right)
\frac  { \ln \left( 10 \right)}{a}
  dM \quad.
\label{equation_mia}
\end {eqnarray}
The parameters that should be deduced from
the  data are $M^*$ ,
$ a$
and  $\Psi^*$.
Table~\ref{datamiamassa} reports the original range in magnitude 
of the astronomical data ,
the selected range adopted for testing purposes ,
the three parameter of our function ,
the $\chi^2$ of the fit and the 
$\chi^2$  of the of the Schechter function
for the five bands of SDSS.
Also here in  the $u^*$ case the  astronomical range and the selected 
range are coincident.
 \begin{table}
 \caption[]{
The full range in magnitudes , 
the selected range in magnitudes ,
the parameters of  our  mass-luminosity function
(\ref{equation_mia})  ,
$\chi^2$ , $AIC$  and $BIC$                     
  of our mass-luminosity function and the Schechter LF 
for the SDSS catalog 
}
 \label{datamiamassa}
 \[
 \begin{array}{lccccc}
 \hline
parameter    &  u^*   &  g^* & r^*  & i & z  \\ \noalign{\smallskip}
 \hline
 \noalign{\smallskip}
full~range~[mags]          
&-22   , -15.8       
&-23.4 , -16.3            
&-24.48 , -16.3            
&-24.48 , -17.2            
&-23.7 , -17.48            
\\ \noalign{\smallskip}
selected~range~[mags]      
&-20   ,  -15.78
&-22 ,  -18.2                      
&-22.94 ,  -18.5                      
&-23.42 ,  -19.3
&-23.7,   -20
\\ \noalign{\smallskip}
a                           
&  0.98           
&  0.95      
&  1.07
&  1.04
&  1.05
\\ \noalign{\smallskip}
M^* [mags]                  
&  -17.27
&  -18.85
&  -19.47
&  -19.98
&  -20.28                 
\\ \noalign{\smallskip}
\Psi^*                      
&   0.05                
&   0.03 
&   0.033
&   0.027
&   0.027               
\\ \noalign{\smallskip}
\chi^2                      
&   552  
&   803 
&   1180 
&   306  
&   475          
\\ \noalign{\smallskip}
AIC, k=3                       
&   558
&   809
&   1186
&   312
&   481
\\ \noalign{\smallskip}
BIC ,k=3                      
&   570
&   821
&   1199
&   325
&   493
\\ \noalign{\smallskip}
\chi^2~Schechter
&   330
&   456
&   1497
&   1863
&   2292                      
\\ \noalign{\smallskip}
AIC~Schechter,k=3                       
&   336
&   462
&   1503
&   1869
&   2298         
\\ \noalign{\smallskip}
BIC~Schechter,k=3                       
&   349
&   474
&   1515
&   1882
&   2310         
\\ \noalign{\smallskip}
 \hline
 \hline
 \end{array}
 \]
 \end {table}
In   the absence of observational  data 
which  represent  the LF ,
 we can generate them 
through  Schechter's parameters, see Table~\ref{parameters};
this is done,  for example, for  the 
CFA Redshift Survey
 ,see~\cite{Marzke_1994b}.
The parameters  of the first  LF ( equation~(\ref{equation_mia_linear})) 
are reported in Table~\ref{para_physical}
where  the requested errors on the
values of luminosity are   the same as    the considered value.
 \begin{table} 
 \caption[]{The parameters of the first LF  \\
            based on  data from CFA Redshift Survey
           ( triplets  generated by the author) } 
 \label{para_physical} 
 \[ 
 \begin{array}{lc} 
 \hline 
~     &   CFA   \\ \noalign{\smallskip}  
 \hline 
 \noalign{\smallskip} 
M^*    [mags]       &  -19  \pm 0.1       \\ \noalign{\smallskip}
\Psi^* [h~Mpc^{-3}] &  0.4  \pm 0.01      \\ \noalign{\smallskip}
 \hline 
 \hline 
 \end{array} 
 \] 
 \end {table}

\subsection{Parameters determination} 

The theoretical  LF  for galaxies  
can be   represented by  an analytical  function
of the type  $\Phi (M^*,\phi^*,p_3)$
where $M^*,\phi^*$ and $p_3$ represent 
the scaling magnitude , the number of galaxies per unit $Mpc$ 
and  a generic third   parameter.
Once the observational data 
are  provided in $n$  triplets made by  absolute magnitude ,
$\phi_{astr}$  ( in  units of number per $h^{-3} Mpc^{-3}$  per mag) 
and  $ \sigma_{\phi} $  ( the error on $\phi$ ) 
we can deduce these three parameters in the   following ways.
\begin{itemize}
\item  A scanning on the presumed values of the parameters that 
       are unknown. The three parameters are those that  minimize
       the  merit function $\chi^2$
       computed as
\begin{equation}
\chi^2 =
\sum_{j=1}^n ( \frac {\phi   - \phi_{astr} } {\sigma_{\phi}})^2
\quad  .
\label{chisquare}
\end{equation}

\item A nonlinear fit  
through the  Levenberg--Marquardt  method ( subroutine
MRQMIN in \cite{press}). In this case the first derivative of the 
LF  with respect to the unknown parameters 
should be provided.
\end{itemize}
Particular attention should be paid to the number of  parameters 
that are unknown : two for the new LF 
as represented by formula~(\ref{equation_mia_linear}) 
, three for the Schechter function (formula~(\ref{equation_schechter_M}))
 and the new mass-LF relationship (formula~(\ref{equation_mia})).
The Akaike information criterion ($AIC$) , see \cite{Akaike1974} ,
is defined as
\begin{equation}
AIC  = 2k - 2  ln(L)
\quad , 
\end {equation}
where $L$ is 
the likelihood  function  and $k$  the number of  free parameters
of the model. 
We assume  a Gaussian distribution for  the errors  
and  the likelihood  function
can be derived  from the $\chi^2$ statistic
$L \propto \exp (- \frac{\chi^2}{2} ) $ 
where  $\chi^2$ has been computed trough equation~(\ref{chisquare}),
see~\cite{Liddle2004},\cite{Godlowski2005}.
Now $AIC$ becomes 
\begin{equation}
AIC  = 2k + \chi^2 
\quad  .
\label{AIC}
\end {equation}

The Bayesian information criterion (BIC), see \cite{Schwarz1978},
is 
\begin{equation}
BIC  = k~ ln(n) - 2  ln(L)
\quad , 
\end {equation}
where $L$ is 
the likelihood  function  , $k$  the number of  free parameters
of the model and  $n$ the number of observations.
 
\subsection{Dependence from the redshift}

The joint distribution in {\it z} (redshift)  and {\it f} (flux)
 for galaxies ,
see formula~(1.104) in~\cite{pad} or formula~(1.117) 
in~\cite{Padmanabhan_III_2002} ,
 is
\begin{equation}
\frac{dN}{d\Omega dz df} =  
4 \pi  \bigl ( \frac {c}{H_0} \bigr )^5    z^4 \Psi (\frac{z^2}{z_{crit}^2})
\label{nfunctionz}  
\quad ,
\end {equation}
where $d\Omega$ , $dz$ and  $ df $ represent the differential of
the solid angle , the redshift and the flux respectively.
The critical value of $z$  ,   $z_{crit}$ , is 
\begin{equation}
 z_{crit}^2 = \frac {H_0^2  L^* } {4 \pi f c_L^2}
\quad .
\end{equation} 
The number of galaxies , $N_s(z,f_{min},f_{max})$  
comprised between a minimum value of flux,
 $f_{min}$,  and  maximum value of flux $f_{max}$ ,
can be computed  through  the following integral 
\begin{equation}
N_S (z) = \int_{f_{min}} ^{f_{max}}
4 \pi  \bigl ( \frac {c}{H_0} \bigr )^5    z^4 \Psi (\frac{z^2}{z_{crit}^2})
df
\label{integrale}
\quad .
\label{integrale_k} 
\end {equation}
This integral does not  have  an analytical solution 
and we    performed  
a numerical integration.
The number of galaxies in {\it z} and {\it f} as given by 
formula~(\ref{nfunctionz})  has a maximum  at  $z=z_{pos-max}$ ,
where 
\begin{equation}
 z_{pos-max} = 1.3798 \times z_{crit}  
\label{maximumia}
\quad ,
\end{equation} 
that can be re-expressed   as
\begin{equation}
 z_{pos-max} =
\frac
{
0.3892\,\sqrt {{10}^{ 0.4\,{\it {\it M_{\sun}}}- 0.4\,{\it M^*}}}{\it
H_0}
}
{
\sqrt {f}{\it c_L}
}
\quad  ,
\label{zmax_k}
\end{equation}
where  $M_{\sun}$ is the reference magnitude 
of the sun at the considered bandpass, 
$H_0$   is the Hubble  constant  and 
$c_L$   is the velocity of the light.
 
\begin{figure}
\begin{center}
\includegraphics[width=10cm]{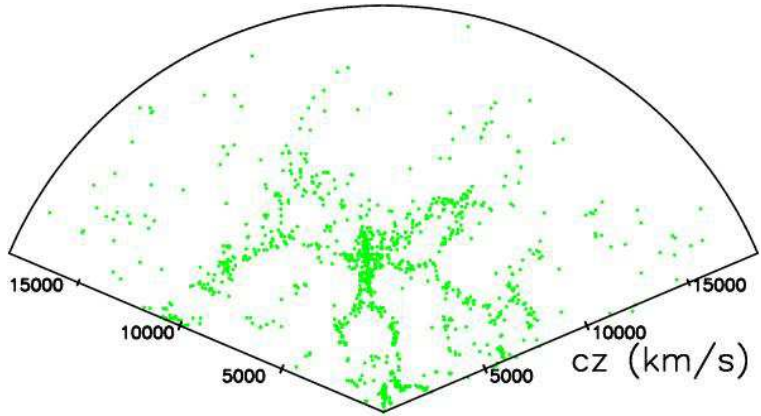}
\end {center}
\caption { Polar plot of   the   real galaxies (green  points)
belonging to the second CFA2 redshift catalogue.}
          \label{simu_cfa}%
    \end{figure}

From the point of view of the astronomical observations 
the second CFA2 redshift   Survey , started in 1984,
produced slices showing that the spatial distribution of galaxies
is not random but distributed on filaments that represent the 2D
projection of 3D bubbles. 
We recall that a slice comprises all the
galaxies with magnitude $m_b~\leq~16.5$ in a strip of $6^{\circ}$
wide and about $130^{\circ}$ long. 
One  such slice (the so
called first CFA strip) is visible at the following address
http://cfa-www.harvard.edu/~huchra/zcat/ 
and  is reported in Figure~\ref{simu_cfa} 
; more details can be
found in~\cite{geller}.  

\begin{figure*}
\begin{center}
\includegraphics[width=10cm]{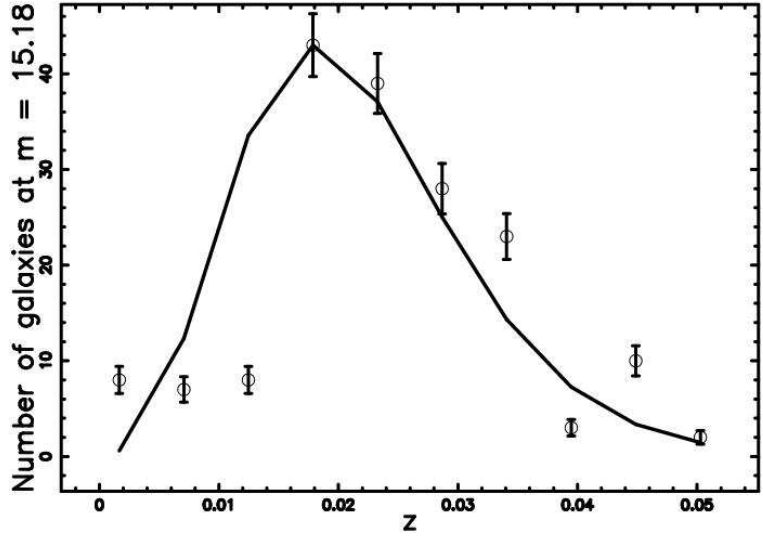}
\end {center}
\caption{
The galaxies  of the second CFA2 redshift catalogue with 
$ 15.08  \leq  mag \leq 15.27 $  or 
$ 48776  \frac {L_{\sun}}{Mpc^2} \leq  
f \leq 58016  \frac {L_{\sun}}{Mpc^2}$
( with $mag$ representing  the 
relative magnitude  used in object selection),
are isolated 
in order to represent a chosen value of $m$ 
and then organized in frequencies versus
heliocentric  redshift
 ,  (empty circles);
the error bar is given by the square root of the frequency.
The maximum in the frequencies of observed galaxies is at  $z=0.02$.
The theoretical curve  generated by 
the z-dependence in the number of galaxies
(formula~(\ref{nfunctionz}) and parameters
as in column CFA  of Table~\ref{para_physical}) 
is drawn  (full line).
}
          \label{maximum_flux}%
    \end{figure*}

Figure~\ref{maximum_flux}
reports the number of  observed  galaxies
of  the second CFA2 redshift catalogue
for a given  magnitude and 
the theoretical curve  as represented by 
formula~(\ref{nfunctionz}).

\begin{figure*}
\begin{center}
\includegraphics[width=10cm]{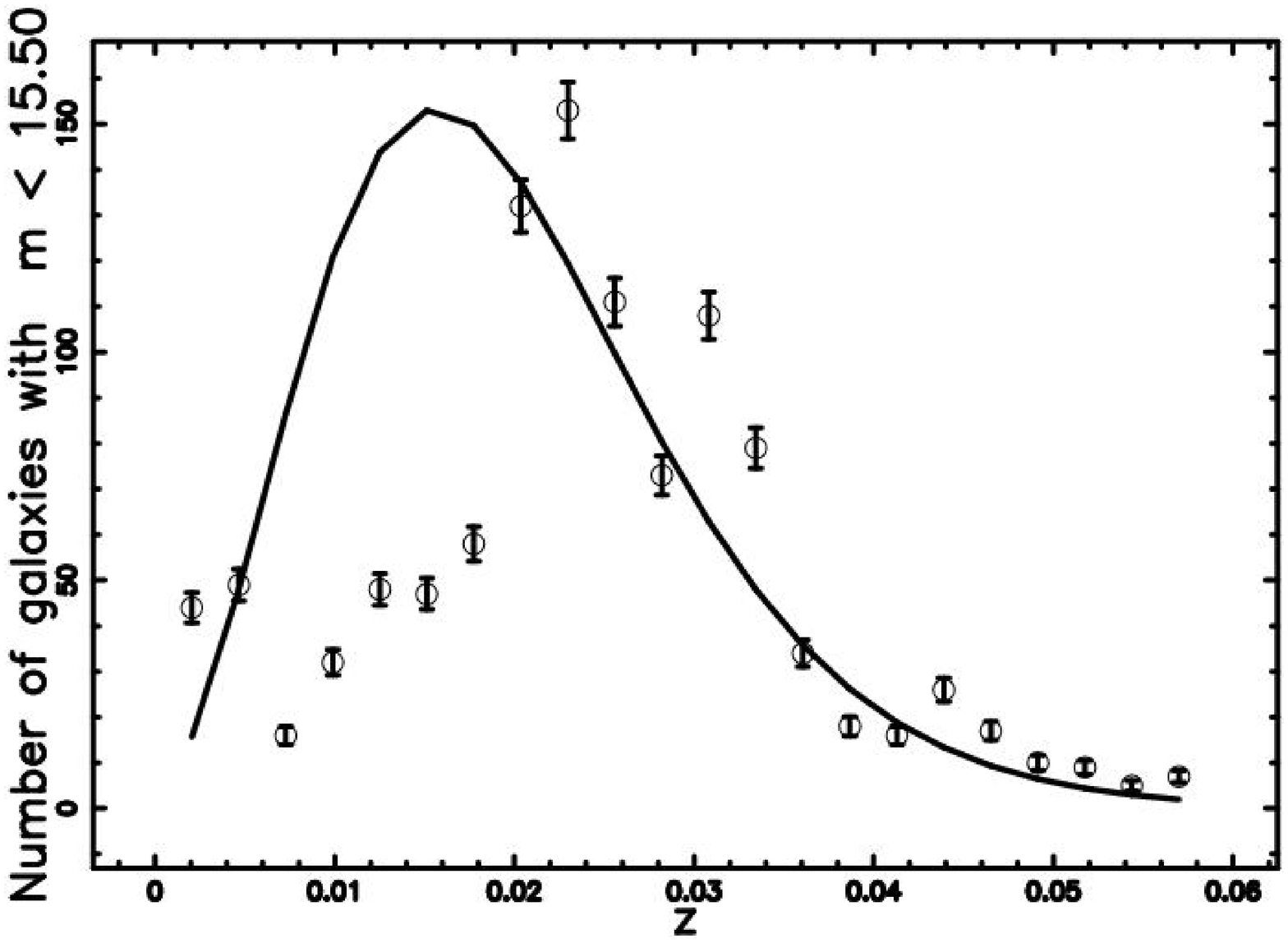}
\end {center}
\caption{
The galaxies  of the the second CFA2 redshift catalogue with 
$ 10.56 \leq mag \leq 15.5 $
or 
$ 39517  \frac {L_{\sun}}{Mpc^2} \leq  
f \leq 3739299 \frac {L_{\sun}}{Mpc^2}$
   ,
are organized in  frequencies versus
  redshift
 ,  (empty stars).
The theoretical curves  generated by
the integral in flux  
(formula~(\ref{integrale})  with parameters
as  in  Table~\ref{para_physical}) 
(full  line) is   drawn.
}
          \label{maximum_flux_all}%
    \end{figure*}

The total number of galaxies of the  second CFA2 redshift catalogue
is reported in Figure~\ref{maximum_flux_all}  as well  as 
the theoretical curve as represented 
by the numerical integration of formula~(\ref{integrale}).

\begin{figure}
\begin{center}
\includegraphics[width=10cm]{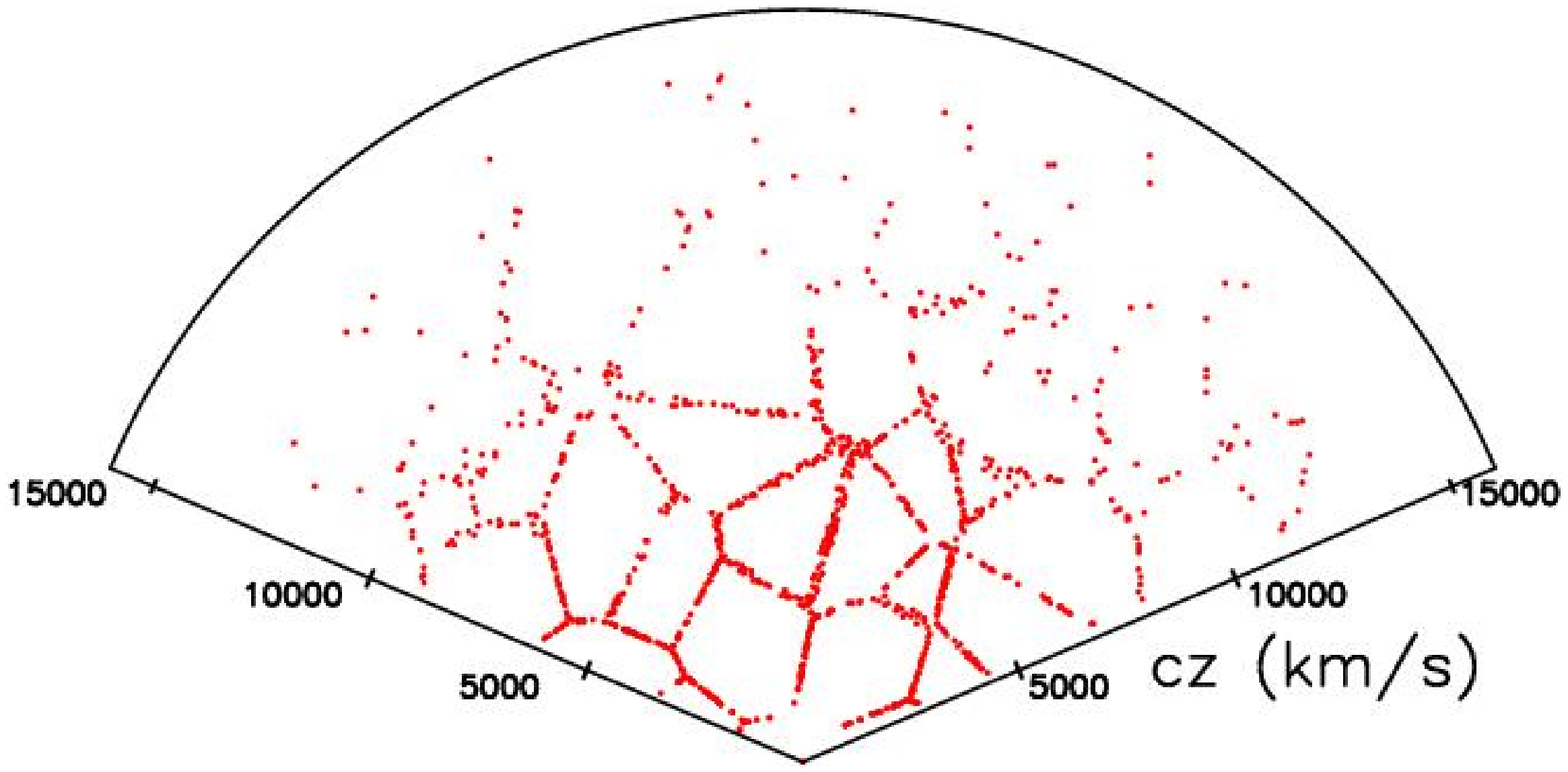}
\end {center}
\caption { 
Polar plot of the little cubes (red points) 
belonging to the simulation.
Parameters  as in Figure~\ref{taglio_mezzo}.
}
          \label{simu_k}%
    \end{figure}

A typical polar plot 
is  reported in Figure~\ref{simu_k}
 once the   number 
of galaxies as a function of $z$  is computed 
through  the numerical integration of formula~(\ref{integrale})
  ; 
it should be compared 
with the observations  , see Figure~\ref{simu_cfa}. 

\section{Summary}

The PDF of the product of two independent random variables
$X$ and $Y$ as represented by two gamma variates with argument 2 
has  been analytically derived.

The mean , the variance and the DF  of this new PDF are computed.
As an application we assumed that the mass of the galaxies 
behaves in the same way  as this new PDF. 
The LF  for galaxies can therefore derived assuming a 
linear or  nonlinear relationship between  mass and luminosity:
in the first case we have  a two parameter LF
and in the second case  a three parameter LF;
recall that the Schechter LF  
for galaxies has three parameters.
The parameter $a$ that characterizes 
the non-linear relationship between mass and luminosity
of the second LF  , see equation~(\ref{equation_mia}) , 
is found to be around 1.

The comparison  between  the two  LF 
for galaxies here derived is   performed  on the SDSS
data and can be done  only  by introducing an upper limit
in magnitude , $M_{lim}$,  in the   five  bands 
 analyzed.
The three  tests of reliability here adopted 
show  that the Schechter function  always has a smaller 
$\chi^2$ , $AIC$ and   $BIC$   with respect to the two new  LF 
for galaxies here derived , 
see Table~\ref{datamia}
and  Table~\ref{datamiamassa}.

The theoretical number of galaxies  as a  function of the red-shift 
presents a maximum that is a function 
of $\alpha$ and $f$   for the 
Schechter  function;
 conversely, when   the  first  new LF  here derived
is considered  , the maximum is a function only of  $f$,
see equation~(\ref{maximumia}).
The first new LF  for galaxies , once   implemented
on a 3D Voronoi slice ,   allow us  to  reproduce 
the large scale structures of our universe.

\end{document}